%% file: main.tex
\documentclass[sigconf, nonacm]{acmart}

\newcommand\vldbdoi{}
\newcommand\vldbpages{}
\newcommand\vldbvolume{16}
\newcommand\vldbissue{10}
\newcommand\vldbyear{2024}

\newcommand\vldbtitle{\shorttitle} 
\newcommand\vldbavailabilityurl{URL_TO_YOUR_ARTIFACTS}
\newcommand\vldbpagestyle{plain} 

\input{header}

\usepackage{bbm, dsfont}

\begin{document}

\title{\sys: A Suite for Responsible Entity Matching \thanks{$^*$Authors contributed equally to this work.}}

\author{Nima Shahbazi$^{* \dagger}$, Mahdi Erfanian$^{* \dagger}$, Abolfazl Asudeh$^\dagger$, Fatemeh Nargesian$^\ddagger$, Divesh Srivastava$^{\dagger\dagger}$}
\affiliation{%
  \institution{$^\dagger$University of Illinois Chicago, $^\ddagger$University of Rochester, $^{\dagger\dagger}$AT\&T Chief Data Office}
  $^\dagger$\{nshahb3, merfan2, asudeh\}@uic.edu, $^\ddagger$fnargesian@rochester.edu, $^{\dagger\dagger}$divesh@research.att.com
}






\begin{abstract}
Entity matching is one of the earliest tasks that occur in the big data pipeline and is alarmingly exposed to unintentional biases that affect the quality of data. Identifying and mitigating the biases that exist in the data or are introduced by the matcher at this stage can contribute to promoting fairness in downstream tasks.
This demonstration showcases \sys, a framework for 1) auditing the output of entity matchers across a wide range of fairness measures and paradigms, 2) providing potential explanations for the underlying reasons for unfairness, and 3) providing resolutions for the unfairness issues through an exploratory process with human-in-the-loop feedback, utilizing an ensemble of matchers.
We aspire for \sys to contribute to the prioritization of fairness as a key consideration in the evaluation of EM pipelines.

\end{abstract}
\maketitle
\submit{\setcounter{page}{1}}
\pagestyle{\vldbpagestyle}
\begingroup\small\noindent\raggedright\textbf{PVLDB Reference Format:}\\
Nima Shahbazi, Mahdi Erfanian, Abolfazl Asudeh, Fatemeh Nargesian, Divesh Srivastava. \vldbtitle. PVLDB, \vldbvolume(\vldbissue): \vldbpages, \vldbyear.\\
\href{https://doi.org/\vldbdoi}{doi:\vldbdoi}
\endgroup
\begingroup
\renewcommand\thefootnote{}\footnote{\noindent
This work is licensed under the Creative Commons BY-NC-ND 4.0 International License. Visit \url{https://creativecommons.org/licenses/by-nc-nd/4.0/} to view a copy of this license. For any use beyond those covered by this license, obtain permission by emailing \href{mailto:info@vldb.org}{info@vldb.org}. Copyright is held by the owner/author(s). Publication rights licensed to the VLDB Endowment. \\
\raggedright Proceedings of the VLDB Endowment, Vol. \vldbvolume, No. \vldbissue\ %
ISSN 2150-8097. \\
\href{https://doi.org/\vldbdoi}{doi:\vldbdoi} \\
}\addtocounter{footnote}{-1}\endgroup

\ifdefempty{\vldbavailabilityurl}{}{
\begingroup\small\noindent\raggedright\textbf{PVLDB Artifact Availability:}
The source code, demonstration video, and other artifacts are available at:  https://github.com/UIC-InDeXLab/FairEMDemo.
\endgroup
}

\input{introduction}
\input{system_overview}
\input{demo}
\input{conclusion}
\section*{Acknowledgement}
This work was supported in part by NSF grants  2107290 and 2348919.

\balance
\bibliographystyle{ACM-Reference-Format}
\bibliography{ref}

\end{document}

%% file: header.tex
\usepackage{flushend}
\usepackage[a-1b]{pdfx}   
\usepackage{balance}

\usepackage{color}
\usepackage[labelfont=bf]{caption}
\usepackage{subcaption}
\usepackage{hyperref}
\usepackage{makecell}
\usepackage{fixltx2e}
\usepackage{enumitem}
\usepackage{rotating,multirow}
\usepackage{etoolbox,xspace}
\usepackage{algorithmicx}
\usepackage{algorithm}
\usepackage{fontawesome}
\usepackage{algpseudocode}

\algtext*{EndWhile}
\algtext*{EndIf}
\algtext*{EndFunction}
\algtext*{EndFor}
\algrenewcommand\algorithmicrequire{\textbf{Input:}}
\algrenewcommand\algorithmicensure{\textbf{Output:}}

\usepackage{amsthm}

\usepackage{tabularx}
\usepackage{blindtext}
\usepackage{ragged2e}  
\usepackage{booktabs}  
\usepackage{textcomp}
\usepackage{url}
\usepackage{comment}
\usepackage{enumitem}
\usepackage{nicematrix}
\usepackage[simplified]{pgf-umlcd}
\usepackage{tikz}
\usepackage{ifthen}
\usepackage{diagbox}

\newcolumntype{Y}{>{\RaggedRight\arraybackslash}X}

\usepackage{booktabs}
\usepackage{xspace}

\newenvironment{proof}{\paragraph{Proof:}}{\hfill$\square$}
\newcounter{example}
\renewcommand{\theexample}{\arabic{example}}

\newcommand{\squishlist}{
 \begin{list}{$\bullet$}
  { \setlength{\itemsep}{0pt}
     \setlength{\parsep}{1pt}
     \setlength{\topsep}{1pt}
     \setlength{\partopsep}{0pt}
     \setlength{\leftmargin}{1em}
     \setlength{\labelwidth}{1em}
     \setlength{\labelsep}{0.5em} } }

\newcommand{\squishend}{
  \end{list}
}

\definecolor{americanrose}{rgb}{1.0, 0.01, 0.24}
\definecolor{airforceblue}{rgb}{0.36, 0.54, 0.66}
\definecolor{ao(english)}{rgb}{0.0, 0.5, 0.0}
\definecolor{ao}{rgb}{0.0, 0.0, 1.0}

\newcommand{\rev}[1]{\textcolor{black}{#1}}
\newcommand{\eat}[1]{}

\newcommand{\submit}[1]{}

\usepackage{xspace}
\newcommand{\stitle}[1]{\vspace{1mm}\noindent{\bf #1:}}

\newcommand{\gee}{\mathbf{g}}

\newcommand{\at}[1]{{\tt \small #1}\xspace}
\renewcommand{\marginpar}[2][]{}

\newcommand{\sys}{\textsc{FairEM360}\xspace}

%% file: introduction.tex
\vspace{-4mm}
\section{Introduction}\label{sec:intro}

Entity matching (EM) is the task of identifying records from one or more data sources that refer to the same real-world entity. EM is a critical step across a wide array of socially sensitive domains such as healthcare, security, HR, elections, and e-commerce. 
Recent studies~\cite{shahbazi2023through, shahbazi2024fairness} have highlighted the significance of evaluating EM tasks with fairness considerations and the potential severity of the consequences if overlooked since there is no single matcher that consistently outperforms all others. 
Certain data properties, such as heterogeneity, quality, inherent similarities among groups, and representation skews, along with the choice of entity matcher may encode unintentional biases towards certain groups resulting in systematic disparate impact. That is, records from some groups may match at a significantly lower/higher rate than records from other groups, with real-world consequences such as under/overestimating the prevalence of certain demographic groups.
In this situation, where results vary based on data, matcher, matching criteria, and fairness measurement, it is necessary to compare the results of different matchers side-by-side, evaluate their advantages and disadvantages, and opt for the most fitting results for the task at hand.

To this end, we propose \sys, a framework for facilitating responsible EM. Our goal is to assist practitioners in finding answers to the following questions:
\begin{itemize}[leftmargin=*]
    \item Does a matcher demonstrate (un)fairness towards a particular group of interest concerning a specific definition of fairness?
    \item Can we find possible explanations for a matcher exhibiting unfairness towards a specific group?
    \item Considering an ensemble of matchers, what strategy yields a desirable trade-off between fairness and matching performance?
\end{itemize}

\sys incorporates a comprehensive set of group fairness definitions tailored for {\em single} and {\em pairwise} fairness metrics specifically designed for EM audits~\cite{shahbazi2023through}.
\sys also features a ready-to-use ensemble of {\em 10 matchers} for performing matching tasks.
Furthermore, \sys offers a diverse range of subgroup-based, measure-based, 
and group-representation based 
explanations that shed light on the reasoning behind matcher's unfair behavior. 
Finally, through an exploratory process with human-in-the-loop feedback that navigates the combinatorial space of matchers, and fairness measures, \sys is directed toward satisfying the fairness constraints required by the user while satisfying a minimum matching performance for various groups.
To the best of our knowledge, {\em \sys stands out as the first system for responsible EM}. \rev{The closest existing system to ours is \textsc{FairER}~\cite{fanourakis2023fairer}, a framework for fair and explainable entity resolution. However, their definition of fairness differs from ours, and they only provide explanations on the matching outcomes, not on the unfairness. Similarly, systems such as \textsc{LEMON}~\cite{barlaug2022lemon}, \textsc{WYM}~\cite{baraldi2023intrinsically}, and \textsc{ExplainER}~\cite{ebaid2019explainer} focus on the explainability of the matching task. Fairness in EM has been studied in various settings: the impact of tuning matching thresholds to achieve fairness ~\cite{moslemi2024threshold}, group-based training for different ethnicities to improve both accuracy and fairness in SVM-based ER~\cite{makri2022towards}, AUC-based fairness for EM/ER tasks and resolving the unfairness through a data augmentation \cite{nilforoushan2022entity}, an experimental analysis of fairness in EM~\cite{shahbazi2023through} and promoting fairness in the data preparation step in EM~\cite{shahbazi2024fairness}.}

%% file: system_overview.tex
\begin{figure}[!t]\label{fig:arch}
    \centering
    \includegraphics[width=\columnwidth]{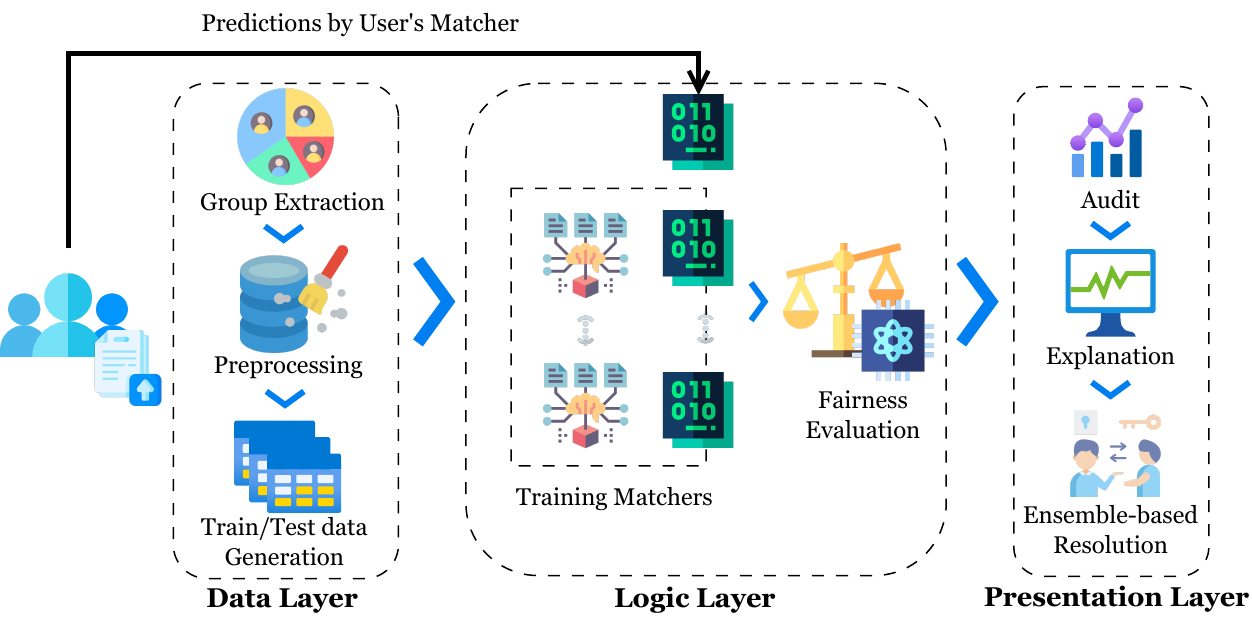}
    \vspace{-8.5mm}
    \caption{\sys Architecture}
    \label{fig:arch}
    \vspace{-6.5mm}
\end{figure}

\vspace{-3mm}
\section{System Architecture}
\vspace{-1mm}

\sys adopts a three-layer architecture: 1) data, 2) logic, and, 3) presentation (Figure~\ref{fig:arch}). In the remainder of this section, we will provide an overview of each layer.

\vspace{-3mm}
\subsection{Data Layer}
The data layer serves as the initial stage when users load their dataset into \sys. This layer fulfills two primary tasks:

\stitle{Group Extraction} The first step in auditing an entity matcher for fairness is identifying meaningful groups/subgroups (e.g., \at{white-male}, \at{black-female}, etc.) from the sensitive attributes (e.g., \at{race}, \at{sex}), based on which the matcher should be audited. Sensitive attributes are attributes for which a matcher is likely to exhibit bias. \rev{Sensitive attributes are determined by a human expert.} Depending on the type, cardinality, and number of sensitive attributes, \sys navigates the space of all possible (sub)groups.
It unifies the group representations using one-hot encoding, which enables representing the subgroups of individual entities and pairs of entities across various settings such as binary, non-binary, and setwise sensitive attributes.
Entity encodings are the output of the data layer and will be passed as an input to the logic layer, where the fairness of a matcher is investigated for a group.

\vspace{-1mm}
\stitle{Preprocessing} If the user opts to utilize the integrated matchers in their evaluation, the data layer will be responsible for preprocessing the input datasets to ensure their compatibility with the matchers, as well as splitting the data into test, train, and validation sets. \rev{Our framework aligns with the format of established benchmark datasets, such as Magellan~\cite{konda2018magellan} and WDC~\cite{primpeli2019wdc}.} The datasets are passed to the logic layer to be trained.

\vspace{-3mm}
\subsection{ Logic Layer}
The actual evaluation of the output from an EM task occurs in the logic layer. The input to the logic layer is a workload of tuples each consisting of left and right entity groups (extracted in the data layer), the matcher prediction, and the matching ground truth. A workload is a test set of tuples for evaluating an entity matcher. Given the pairwise nature of EM tasks, {\em single fairness} or {\em pairwise fairness} paradigms can be used to audit matchers for fairness. 
In single fairness, the performance of a matcher is evaluated for one group, matching at least one of the tuples in a pair. In pairwise fairness though, a pair of groups is considered for evaluation.
Given a group of interest, the logic layer summarizes the workload into a confusion matrix, which is used in computing various fairness measures. Next, proper fairness measures are used to evaluate the EM task depending on the context of the task, the fairness paradigm of interest, etc. \sys offers a set of {\em 5 group fairness definitions} specific to EM that users can use to evaluate their tasks. Finally, each group's unfairness is calculated using {\em subtraction-based} or {\em division-based} notations of disparity and if the unfairness exceeds the fairness threshold specified by the user, the matcher is considered unfair for that group.
In the following, we provide more details on fairness definitions and unfairness measurement: 

The input to the logic layer is a workload $w$ of $n$ tuples each having a correspondence $t=(e_i, e_j, h, y)$, where $h$ is a binary variable indicating the result of entity matching ({\em match} or {\em non-match}) for entities with encodings $e_i$ and $e_j$, and $y$ is a binary variable indicating the ground-truth for matching. 
A workload is defined as a test set of tuples for evaluating an entity matcher. 

\stitle{Fairness Paradigms}
Given the pairwise nature of EM tasks, there are two paradigms to audit matchers for fairness.
In the {\em single fairness}, the performance of a matcher is evaluated for one subgroup $s$ against either entity in a pair.
Given a correspondence $(e_i, e_j, h, y)$ and a subgroup $s$ of interest, \sys considers the correspondence legitimate, if either $e_i$ or $e_j$ belong to subgroups $s$. 
In {\em pairwise fairness}, the performance of a matcher is evaluated for a pair of subgroups $s,s^\prime$ against both entities in a pair. 
Given a correspondence $(e_i, e_j, h, y)$ and a pair of sub-groups $s,s^\prime$ of interest, \sys considers the correspondence legitimate, if either $e_i$ or $e_j$ belong to subgroups $c$. if $e_i$ belongs to $s$ and $e_j$ belongs to $s^\prime$, or vice versa.

Given a subgroup of interest, the logic layer summarizes the workload $w$ into a confusion matrix, which is later used in computing various fairness measures.
When auditing with single and pairwise fairness, each correspondence $(e_i,e_j,m,y)$ corresponds either to True Positives (TPs), False Positives (FPs), False Negatives (FNs), or True Negatives (TNs). 
Note that the result is counted both for the group(s) of $e_i$ and the group(s) of $e_j$.
This is unlike regular classification where every row 
corresponds to one entity and hence is counted once.

\stitle{Fairness Measures}
Depending on the context of an EM task at hand, proper fairness measures should be employed.
Besides, a major difference between EM and regular classification tasks is that the input to EM tasks is a pair of records.
Due to its pairwise matching nature, {\em class imbalance} is a distinguishing property of EM, compared to regular classification tasks. In these settings, the fairness measure for successfully discovering these events is {\em Positive Predictive Value Parity (PPVP)}. Another important measure in this context is {\em True Positive Rate Parity (TPRP)}, a.k.a {\em Equal Opportunity}, which focuses on correct match predictions among the (rare) true matches. For further information regarding the fairness measures in the context of EM, refer to ~\cite{shahbazi2023through}.

\stitle{Measuring Unfairness}
Consider a fairness notion and a subgroup $g_i\in \gee$. In a perfect situation, the matcher should satisfy the parity (equality) between two probabilities in the following form:
\begin{equation}
    \label{equation:general}
\forall g_i\in \gee, Pr(\alpha~\vert~ \beta,g_i) = Pr(\alpha~\vert~ \beta)
\end{equation}
where $\alpha$ and $\beta$ are specified by the fairness measure.
For example, for Positive Predictive Parity, $\alpha$ is $y=`M\textrm'$ and $\beta$ is $h(e,e^\prime)=`M\textrm'$.

On the other hand, due to the trade-offs between different fairness notions and the impossibilities theorems, it is often not possible to satisfy complete parity on all fairness measures.
As a result, considering a threshold value (e.g. the 20\% rule suggests the threshold as $0.2$), the objective is to make sure that {\em disparity} (a.k.a {\em unfairness}) is less than the threshold.
Given a fairness notion and a subgroup $g_i\in \gee$, the disparity can be computed using subtraction, as follows:
\begin{equation}
    \label{equation:ds}
F^{(s)}_{\alpha,\beta}(g_i) = \max \Big(0~,~ Pr(\alpha~\vert~ \beta) - Pr(\alpha~\vert~ \beta,g_i)\Big)
\end{equation}

Alternatively, given a fairness notion and a subgroup $g_i\in \gee$, the disparity can be computed using division, as follows:
\begin{equation}
    \label{equation:dd}
F^{(d)}_{\alpha,\beta}(g_i) = \max \Big(0~,1 - \frac{Pr(\alpha~\vert~ \beta,g_i)}{Pr(\alpha~\vert~ \beta)}\Big)
\end{equation}
In \sys, we provide the choice to utilize either {\em subtraction-based} and {\em division-based} notation of disparity.

\vspace{-.5mm}
\stitle{Training Matchers} 
If the user has chosen to use the integrated matchers in their evaluation, this layer initially trains the matchers before passing them to the evaluation component for auditing. We have integrated 10 ML-based EM systems into this component. Broadly speaking, these matchers can be categorized into one of two groups of {\em non-neural} or {\em neural} including:

\vspace{-1.5mm}
\begin{itemize}[leftmargin=*]
    \item {\em Non-neural.} {\small{ {\at{DTMatcher}} \cite{konda2018magellan}}, {\at{SVMMatcher} \cite{konda2018magellan}}, {\at{RFMatcher} \cite{konda2018magellan}}, {\at{LogRegM-} \at{atcher} \cite{konda2018magellan}}, {\at{LinRegMatcher} \cite{konda2018magellan}}, {\at{NBMatcher} \cite{konda2018magellan}}.}
    \item {\em Neural.} {\small{{\at{DeepMatcher} \cite{mudgal2018deep}}, {\at{Ditto} \cite{li2020deep}}, {\at{HierMatcher} \cite{fu2021hierarchical}}, {\at{MCAN} \cite{zhang2020multi}}}}.
\end{itemize}

 \vspace{-1mm}
 The containerized design of the matcher integration component has greatly facilitated its extension. Practitioners can easily integrate their EM systems into \sys by containerizing them and implementing a wrapper to preprocess the input datasets into their required format.

\vspace{-3mm}
\subsection{Presentation Layer}
The presentation layer analyzes the results generated by the logic layer from evaluating fairness measures on (sub)groups of interest.  More concretely, the input to the presentation layer is the disparity values of each group in case of single fairness or each group pair in case of pairwise fairness, for all applicable measures. The presentation layer has three main components:

\vspace{-1mm}
\stitle{Audit}
The audit component presents the fairness evaluation results to the user. For every fairness paradigm and relevant fairness measure, \sys illustrates the groups to which a matcher has exhibited unfairness and quantifies the extent of this disparity. 

\rev{\stitle{Multiple-workload Analysis} We also consider the scenario where multiple instances of test data a.k.a workloads are available from a matcher, i.e. different test data may become available at different times or different samples from the underlying test data distribution may be available. 
In this case, the logic layer evaluates workloads separately. 
For $k$ workloads, the input to the presentation layer is a population of $k$ disparity values for each group and measure combination. 
A population for group $g$ and measure $m$ includes the disparity of $g$ in every workload w.r.t. $m$. 
To assess the fairness of a matcher on group $g$  using measure $m$ and $k$ workloads, we employ the standard \textit{hypothesis testing}. For a group $g$ and measure $m$, the fairness hypothesis testing considers the \textit{null} hypothesis that the matcher is fair on $g$ and the \textit{alternative} hypothesis that the matcher is unfair on $g$.  For more than one workload, \sys chooses the appropriate $z$-test statistics. Next, the test statistics and corresponding $p$-value are computed as the probability of getting the observed test statistic or something more extreme when the null hypothesis is true. Finally, given a significance level $\alpha$, the null hypothesis in favor of the alternative is rejected if $\alpha\leq p$-value, or not if $\alpha>p$-value. 
If a user chooses to perform multiple-workload analysis on a single provided test data set, \sys generates $k$ workloads by random sampling with replacement from the data set. The goal of this analysis would be to verify whether an unfair group happened by chance or is indeed repeatable.}

\vspace{-1mm}
\stitle{Explanation}
After a matcher is audited based on fairness measures and groups, and unfairness are identified, \sys reassures to offer additional insights explaining the unfairness towards a group. The explanations provided by \sys fall under the category of {\em Local Model-agnostic Methods}~\cite{molnar2020interpretable}, where given an unfairness measure and a group for which the model has been unfair, the goal is to provide (local) explanations for the queried (measure, group).

The presentation layer determines whether a matcher w.r.t. a measure/group. 
To allow users to explore potential explanations for unfairness, \sys provides four perspectives:

\begin{itemize}[leftmargin=*]
    \item {\em Subgroup-based Explanation.} A matcher may be unfair on a group (e.g., \at{female}) because it performs poorly on more granular subgroups (e.g., \at{black-female}). 
    Navigating the subgroup hierarchy of a matcher from an unfair group to its subgroups while considering the matcher's performance on subgroups allows us to identify the subgroups that may be the source of unfairness. 
    Assuming sufficient data exists for the unfair group and its granular subgroups in the dataset, disparity analysis of these granular subgroups over various measures allows the user to gain more insights into the unfairness of the original unfair group.
    
    \item {\em Measure-based Explanation.}
    The measure-based explanation describes the unfairness of a matcher subject to a specific group in terms of the group's confusion matrix. 
    For example, the low accuracy for a specific group can be due to the high false-positive rate for that group.
    This is a common practice in analyzing matcher performance holistically. 
    Measure-based explanations offer insights into the fairness definitions, indicating 
    what factors have contributed to the emergence of such unfairness.
    
    \item {\em Group-representation based Explanation.} Responsible training in EM techniques requires access to unbiased data with proper representation of different groups and possible cases~\cite{shahbazi2023representation}.
    Over/under-representation of different groups can bias the models in favor of some of the groups, making the model unfair. In particular, given the class imbalance nature of EM tasks, it is important to ensure proper representatives from different groups in both (match/unmatch) classes. This group of explanations analyzes the representation of each group within the dataset and conditioned on the matching ground truths.
    \item {\em Example-based Explanation.}
    For each group that exhibits unfairness according to a particular fairness measure, \sys randomly selects a small sample for user review. These examples serve to provide the user with deeper insights into the factors (e.g., name similarities) contributing to higher error rates within this specific group. Such factors may include inherent similarities among certain groups or matchers assigning greater importance to specific features, among others.

\end{itemize}

\vspace{-2mm}
\stitle{Ensemble-based Resolutions} The primary goal of this component is to resolve the existing unfairness by
assigning various matchers for various groups (e.g., \at{HierMatcher} for \at{white} and \at{SVM-based} for \at{black} individuals).
Let $E:\mathcal{G}\rightarrow \mathcal{M}$ be an ensemble, where each groups $g\in\mathcal{G}$ is assigned the model $E(g)$.
One strategy is to use the best-performing matcher for each group $g\in\mathcal{G}$. That is $E(g) = \mbox{argmax}_{M\in\mathcal{M}} \big(A_M(g)\big)$, where $A_M(g)$ is the performance of the model $M$ for group $g$. This approach, while being optimal for each group may not be fair, as the best-performing models for various groups may not provide an equal performance between all groups.
Alternatively, optimizing for fairness, one can find an assignment of models with minimum performance disparity across different groups (i.e., max. fairness). This approach, however, may miss to assign good-performing models for some groups.
Therefore, instead of sticking to one strategy, \sys enables optimization based on both strategies.
To do so, it considers two criteria: (a) the worst performance $A$ for a group, where the model performance for all groups is at least $A$; and (b) unfairness $F$.

Consider the set of all possible assignment of models to the groups. Note that for $k$ groups and $m$ models, there are $k^m$ such assignments $\{E_1, E_2,\cdots,E_{k^m}\}$.
Let $A_i$ and $F_i$ be performance and fairness values for each assignment $E_i$.
\sys views each $E_i$ as a point $\langle F_i, A_i\rangle$ in the fairness-performance space. It then shows the {\em Pareto-frontier} of the points, i.e., set of non-dominated assignments, to the user in order to visually explore the trade-offs of fairness and performance, and to select a non-dominated assignment $E_i$.


\vspace{-3mm}
\subsection{Implementation Details}

The \sys is a web-application with front-end and back-end modules.
The back-end is responsible for data layer and logic layer operations, while the front-end handles visualization tasks. The back-end is implemented using \textsc{FastAPI} to maximize adaptability and scalability. Employing abstraction and object-oriented programming, the architecture facilitates seamless integration of new matchers with fairness measures. Communication with embedded matchers is facilitated through \textsc{Docker} containers, enabling each matcher to operate with its own dependencies. This design enables the system to accommodate multiple matchers concurrently. Introducing a new matcher simply requires containerizing it and implementing designated abstract methods for preprocessing, running and extracting final scores.
The front-end is implemented using \textsc{ReactJS}, a popular JavaScript library that facilitates the development of modular and reusable user interfaces. \textsc{ReactJS} employs a component-based architecture, where state components manage data within the component hierarchy. This approach enables a fully dynamic and interactive system, capable of adapting seamlessly to user interactions.
Furthermore, the back-end and front-end employ a Representational State Transfer (RESTful) API for communication, thereby establishing independence between the system's logic and visualization components. This decoupling facilitates independent modifications to either part without cascading effects on the other. 

%% file: demo.tex
\begin{figure*}[!tb] 
    \centering
    \begin{minipage}[t]{0.33\linewidth}
        \centering
        \includegraphics[width=\textwidth]{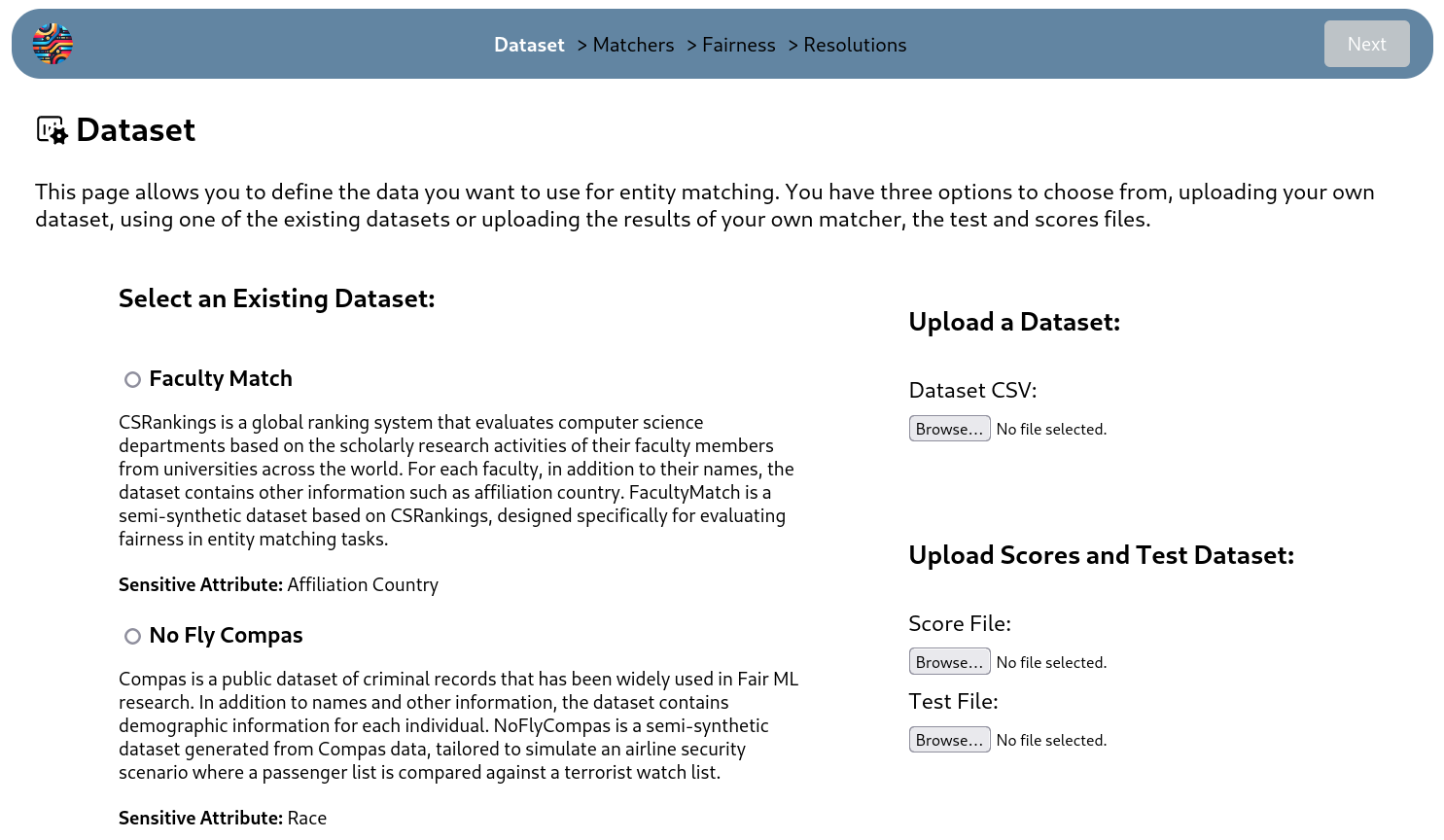}
        \vspace{-7mm}\caption{data import step.}
        \label{fig:dataset}
    \end{minipage}
    \hfill
    \begin{minipage}[t]{0.33\linewidth}
        \centering
        \includegraphics[width=\textwidth]{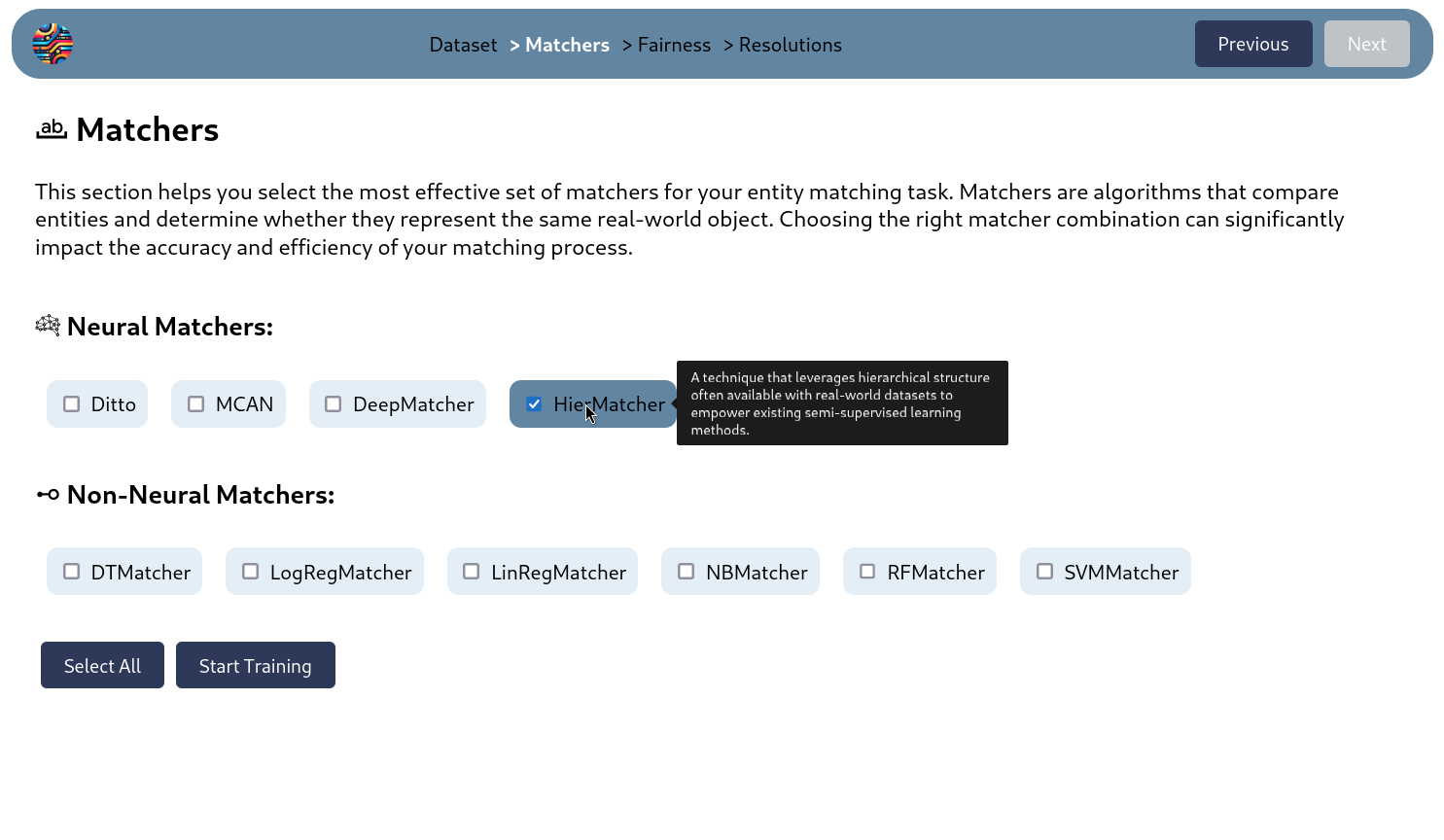}
        \vspace{-7mm}\caption{matcher selection step.}
        \label{fig:matcher}
    \end{minipage}
    \hfill
    \begin{minipage}[t]{0.33\linewidth}
        \centering
        \includegraphics[width=\textwidth]{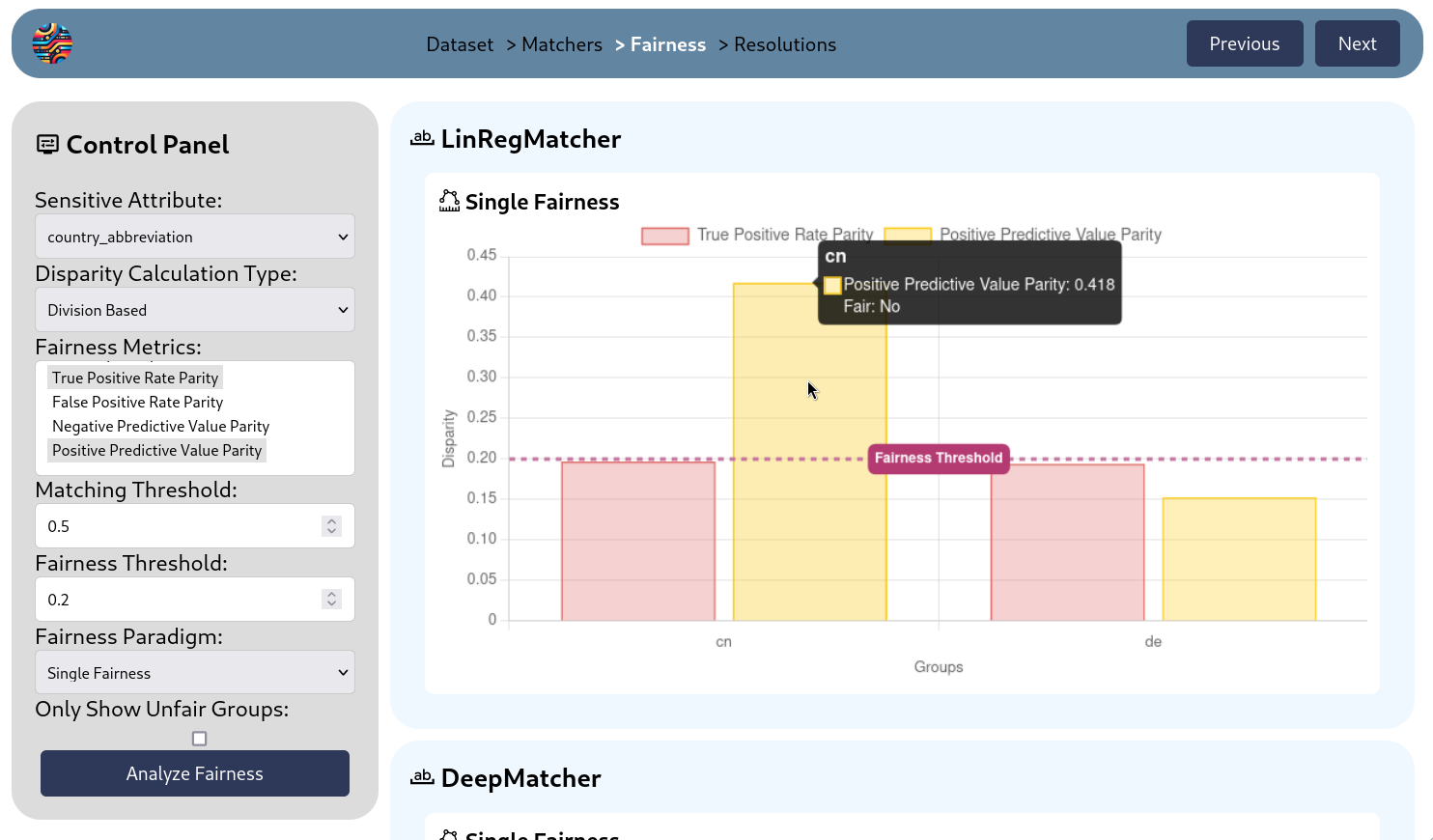}
        \vspace{-7mm}\caption{audit step. evaluation criteria (left) and audit results (right)}
        \label{fig:eval}
    \end{minipage}
    \hfill
    \begin{minipage}[t]{0.33\linewidth}
        \centering
        \includegraphics[width=\textwidth]{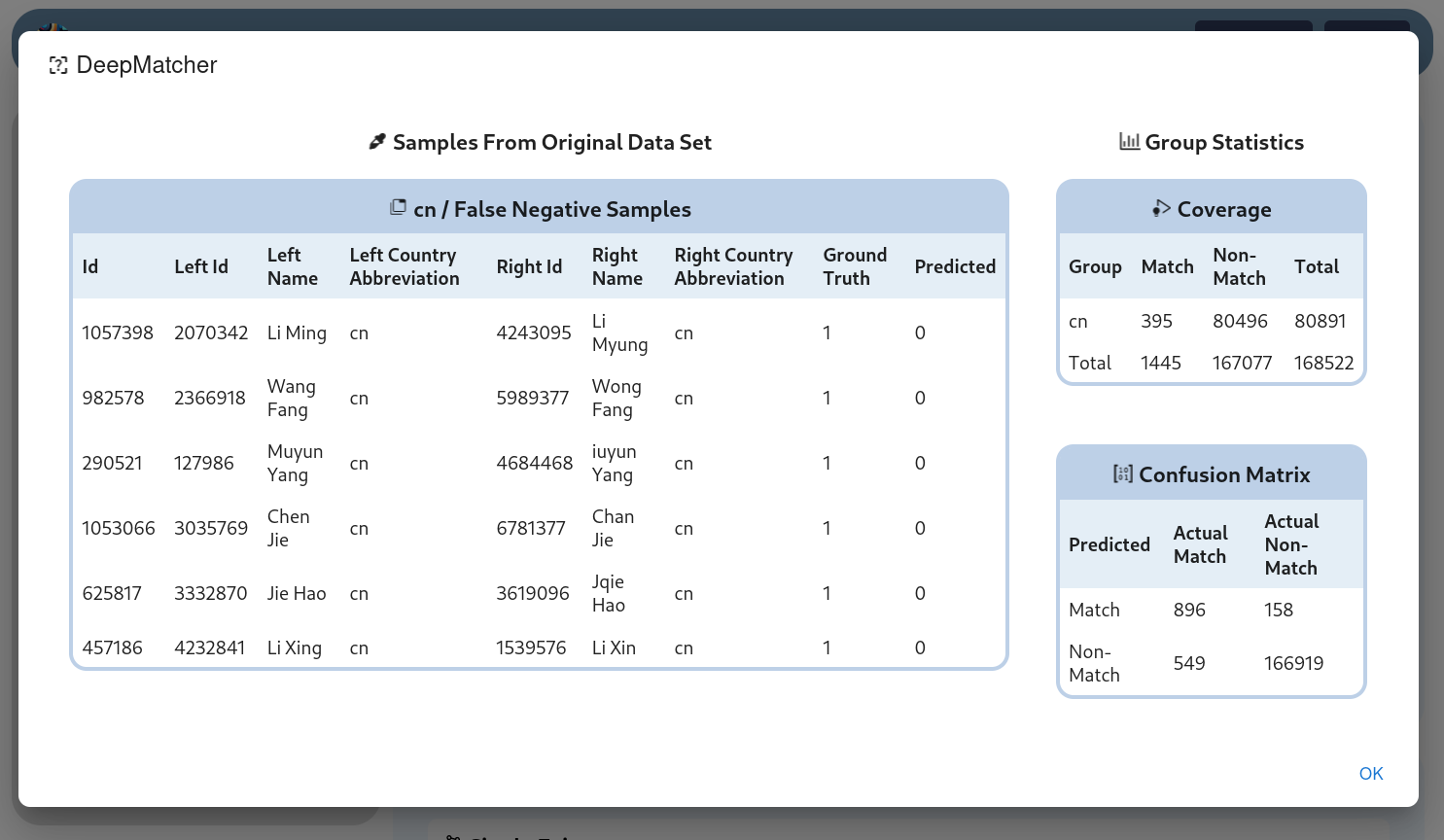}
        \vspace{-7mm}\caption{unfairness explanations for the  observed in the {\sf cn} group w.r.t. TPRP.}
        \label{fig:explanation}
    \end{minipage}
    \hfill
    \begin{minipage}[t]{0.33\linewidth}
        \centering
        \includegraphics[width=\textwidth]{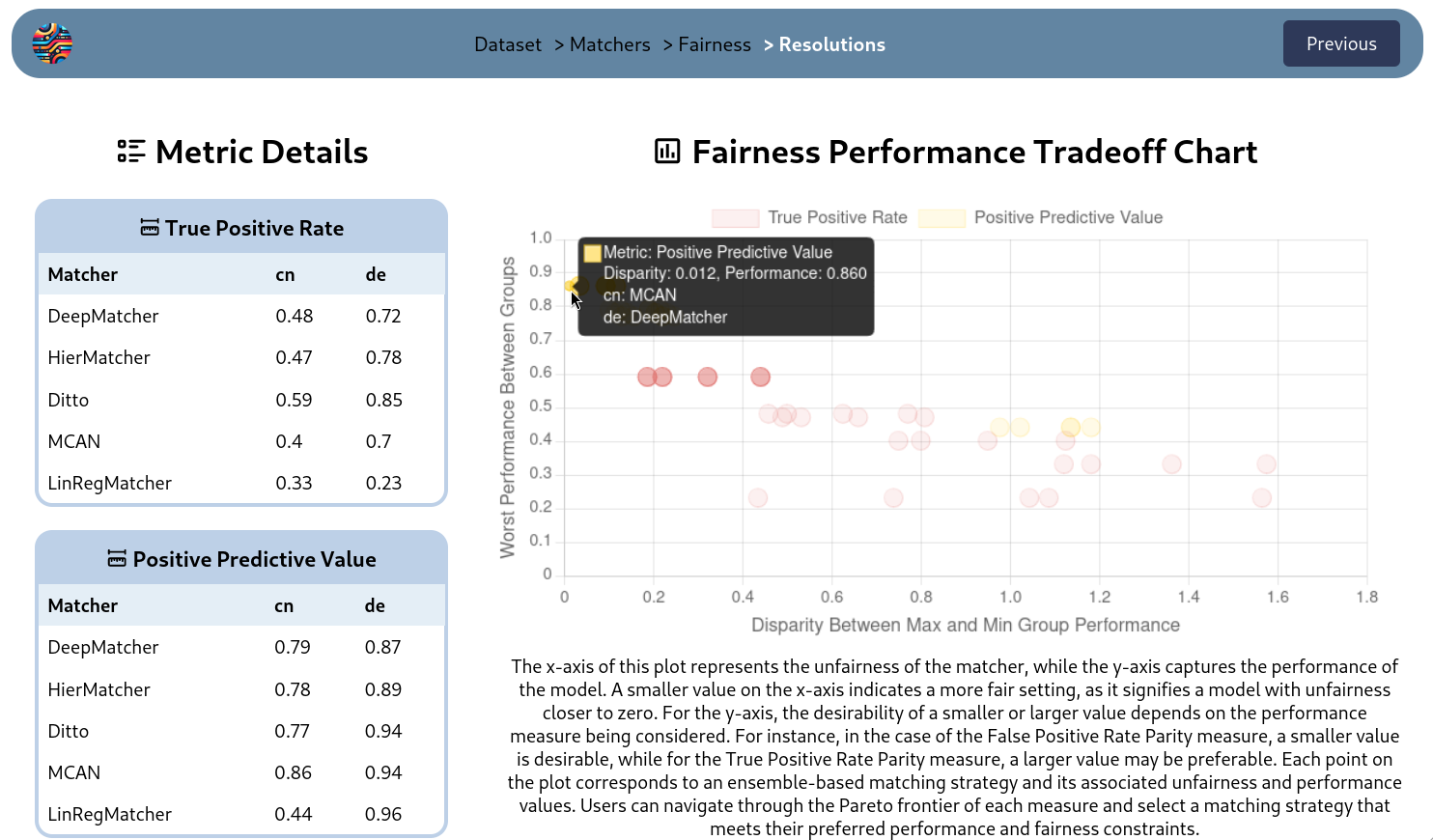}
        \vspace{-7mm}\caption{ensemble-based resolutions.}
        \label{fig:resolution}
    \end{minipage}
    \hfill
    \begin{minipage}[t]{0.33\linewidth}
        \centering
        \includegraphics[width=\textwidth]{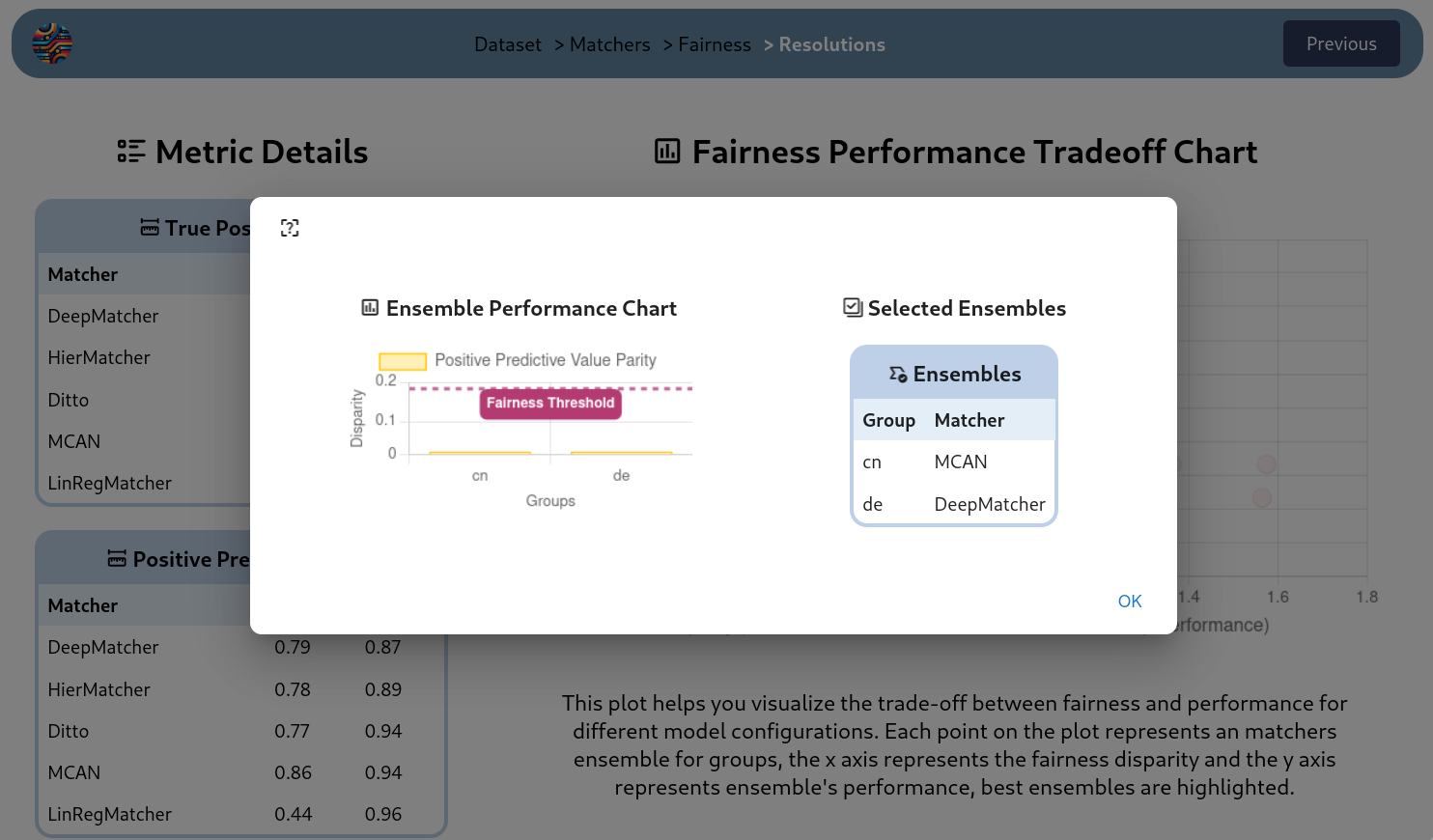}
        \vspace{-7mm}\caption{matching strategy based on ensemble-based resolutions.}
        \label{fig:strategy}
    \end{minipage}
    \vspace{-5mm}
\end{figure*}

\vspace{-3mm}
\section{Demonstration Scenarios}
\vspace{-1mm}

In this demonstration, our objective is to illustrate how \sys can aid practitioners in auditing their EM tasks. The landing page of \sys greets users with an introduction and offers essential details about the system and how it operates. Following that, a standard workflow within \sys involves four key steps:

\stitle{Step 1: Data Import} In this step, users are asked to import their datasets into the system. At a high level, \sys performs two distinct tasks: (i) {\em Matching-and-Evaluation} and (ii) {\em Evaluation-Only}. 
In the former task, users upload a dataset to the system and utilize the integrated matchers within the system to conduct the matching process, followed by evaluating the matching outcomes. In the latter task, users have already executed the matching process using their own matcher, and they upload the predictions (in the format of scores) along with the dataset to the system for evaluation. In our demonstrations, we use {\sf FacultyMatch} and {\sf NoFlyCompas} datasets to highlight the capabilities of \sys. Figure~\ref{fig:dataset} illustrates the data import step. The user has selected the {\sf FacultyMatch} dataset and proceeds to the next step.

\stitle{Step 2: Matcher Selection}
Depending on the task of interest, during this step, the user selects a set of matchers to execute the matching task on the input dataset. This is particularly useful when users wish to compare the performance of multiple matchers on their data or assess the performance of their own matcher against the integrated matchers \sys and leverage them to resolve any potential unfairness issues. Users can also access information about each matcher by hovering over its title. Figure~\ref{fig:matcher} illustrates the matcher selection step. 

\stitle{Step 3: Fairness Evaluation}
In this step, the user initially determines how to conduct the audit on the matching results. The initial decision involves selecting the sensitive attribute which assists \sys in the automatic group extraction process. Next, the user must select the unfairness calculation approach (subtraction-based vs. division-based) and specify the fairness measures they wish to employ to evaluate their task. Users can view information about each definition by hovering over the corresponding measure.
Next, the user should specify the matching threshold, a value within the range of $[0,1]$ that determines the cut-off point above which a pair is considered a match. The subsequent variable to be determined is the fairness threshold, which specifies the threshold beyond which the disparity between a specific group and the average value based on a fairness definition is considered unfair. 
Lastly, the user selects the fairness paradigm and decides whether they want to exclusively view the results related to unfair groups.

With the chosen evaluation criteria as outlined above, \sys initiates the audit process and displays the results in the right pane. For each matcher and fairness paradigm, the results are presented as bar charts illustrating the unfairness value for each group alongside each measure. 
Any group whose value surpasses the designated fairness threshold (highlighted by a red line in the plot) is considered unfair. The plots provide a high level of interactivity, enabling users to filter results by clicking on the measure name in the legend or to view details of each group by hovering over individual bars.
Figure~\ref{fig:eval} illustrates an instance of the audit \sys offers where \at{LinRegMatcher} demonstrates unfairness for the {\sf cn}, as its unfairness of 0.418 exceeds the specified fairness threshold of 0.2.

By clicking on each bar, \sys provides insights into various aspects of the group, including its representation (coverage) in the training data, a random sample of problematic pairs associated with the measure and group, investigating the subgroup hierarchy unfairness if applicable (specific to setwise or intersectional sensitive attributes), group's confusion matrix and factors that have contributed to the unfairness, among other details. 
Figure~\ref{fig:explanation} provides an example of the potential explanations for a matcher's unfair behavior towards a group. In this case the matcher has been unfair towards {\sf cn} group w.r.t. {\sf True Positive Rate Parity}. By examining the example-based explanations, one can infer that the matcher is erroneously matching entities from the {\sf cn} group due to inherent similarities present in Chinese names compared to German names. \rev{Note that any dataset with any grouping of data for which we require equal performance of the matcher can be evaluated by \sys. It does not necessarily need to be social data; however, in the demo, we focus only on such datasets. For analysis on non-social benchmark datasets, please refer to~\cite{shahbazi2023through}.}

\vspace{-1mm}
\stitle{Step 4: Ensemble-based Resolution} Having observed an instance of unfairness towards a group, the next logical step is to propose a resolution. Therefore, in the final step, user specifies a group for which they want to resolve the unfairness issues.
From that point, \sys adopts an exploratory approach based on an ensemble of matchers. For each disadvantaged group, an alternative matcher that performs \underline{\em superior} can be selected to carry out the matching task. However, this superiority is contingent upon the user's preference regarding the priority of group accuracy versus fairness. User specifies an accuracy measure they want to optimize the matching task for. Next, \sys presents the user with the settings that yield the best performance while simultaneously achieving fairness. However, as it may not always be feasible to find such a matcher, the user may choose whether they can accept slightly lower yet still acceptable performance while ensuring fairness, or if they prefer to prioritize a more accurate result even if it meets a less strict fairness threshold. \sys facilitates this process by presenting the user fairness/performance trade-off plot highlighting the Pareto frontier. The x-axis of this plot represents the unfairness of the matcher, while the y-axis captures the performance of the model. A smaller value on the x-axis indicates a more fair setting, as it signifies a model with unfairness closer to zero. For the y-axis, the desirability of a smaller or larger value depends on the performance measure being considered. For instance, in the case of the {\sf False Positive Rate Parity} measure, a smaller value is desirable, while for the {\sf True Positive Rate Parity} measure, a larger value may be preferable. 
Each point on the plot corresponds to an ensemble-based matching strategy and its associated unfairness and performance values. Users can navigate through the Pareto frontier of each measure and select a matching strategy that meets their preferred performance and fairness constraints. Finally, by clicking on the desired point, users can view the recommended strategy from our resolution component for matching and observe the audit results for such a strategy. Figure~\ref{fig:resolution} illustrates an example of the resolution step. In this scenario, the user can choose \at{MCAN} matcher, which offers a 0.926 {\sf Positive Predictive Value} and a 0.056 (i.e., 5.6\%) unfairness to perform the matching for the {\sf cn} group. Figure~\ref{fig:strategy} displays the audit results based on this matching strategy, which effectively resolves the unfairness issue.

This iterative process continues until the user is satisfied with the results, and eventually, the user is presented with the final audit results based on their decisions.

%% file: conclusion.tex
\vspace{-3mm}
\section{Conclusion}
\vspace{-1mm}
In this paper, we introduced \sys, a comprehensive suite designed for fairness-aware entity matching. Practitioners can use \sys to conduct audits on their EM tasks, acquire insights into the root causes of unfairness exhibited by matchers, and address these fairness issues using an ensemble-based approach. We hope that this tool serves as a step towards fostering responsible data analytics practices.
\vspace{-3mm}